\documentclass[twocolumn,showpacs,preprintnumbers,amsmath,amssymb]{revtex4}
\usepackage{graphicx}
\usepackage{dcolumn}
\usepackage{bm}

\begin{document}
\title{Structures and Thermodynamics of the Mixed Alkali Alanates}
\author{J. Graetz,Y. Lee$^*$, J.~J. Reilly, S. Park$^*$, T. Vogt$^*$}
 \affiliation{Department of Energy Sciences and Technology\\$^*$Department of Physics\\Brookhaven National Laboratory\\Upton, New York 11973}
\date{\today}
\baselineskip=10pt
\begin{abstract} The thermodynamics and structural properties of the hexahydride alanates (M$_2$M$'$AlH$_6$) with the elpasolite structure have been investigated. A series of mixed alkali alanates
(Na$_2$LiAlH$_6$, K$_2$LiAlH$_6$ and K$_2$NaAlH$_6$) were synthesized and found to reversibly absorb and desorb hydrogen without the need for a catalyst. Pressure-composition isotherms were measured
to investigate the thermodynamics of the absorption and desorption reactions with hydrogen. Isotherms for catalyzed (4 mol\% TiCl$_3$) and uncatalyzed Na$_2$LiAlH$_6$ exhibited an increase in
kinetics, but no change in the bulk thermodynamics with the addition of a dopant.  A structural analysis using synchrotron x-ray diffraction showed that these compounds favor the $Fm\bar{3}m$ space
group with the smaller ion (M$'$) occupying an octahedral site. These results demonstrate that appropriate cation substitutions can be used to stabilize or destabilize the material and may provide
an avenue to improving the unfavorable thermodynamics of a number of materials with promising gravimetric hydrogen densities.
\end{abstract}
\maketitle

\section{Introduction} There is currently much interest in the development of a sustainable hydrogen storage material for mobile applications. The key requirements for any candidate material for
practical on-board hydrogen storage are a high gravimetric hydrogen density and safe, fast and fully reversible hydrogenation near ambient conditions. Conventional metal
hydrides that can readily supply hydrogen at room temperature have storage capacities $< 2$ wt.\%  and cannot satisfy this need. However, a number of complex hydrides have appreciable gravimetric
hydrogen storage capacities, such as the sodium alanates, which reversibly absorb/desorb hydrogen with the addition of a metal dopant \cite{Bogdanovic1997,Bogdanovic2000}. At present,there is
considerable interest in understanding the kinetic enhancements attributed to the transition-metal dopant \cite{Sandrock2002,Kiyobayashi2003,Fichtner2004} and the mechanism by which the dopant makes
the sodium alanates reversible \cite{Gross2000,Sun,Bogdanovic2003,Brinks2004,Graetz2004,Felderhoff2004,Iniguez2004}. The reversible hydrogenation of NaAlH$_4$ occurs in two steps in the presence
of Ti:
\begin{equation}
3\textup{NaAlH}_4 \leftrightarrow \textup{Na}_3\textup{AlH}_6+2\textup{Al}+3\textup{H}_2\textup{ (3.7 wt.\%)}
\label{naalh4}
\end{equation}
\begin{equation}
2\textup{Na}_3\textup{AlH}_6 \leftrightarrow 6\textup{NaH}+2\textup{Al}+3\textup{H}_2\textup{ (1.9 wt.\%)}.
\label{na3alh6}
\end{equation}
While the atomic mechanism is not understood, it appears that Ti resides in an under-coordinated environment at or near the surface of the depleted material \cite{graetz_mrs,Graetz2004}.
Despite the recent attention, sodium aluminum hydride is unlikely to meet the requirements necessary for automotive applications \cite{DOE}. A few alternatives that exhibit higher
hydrogen capacities are LiAlH$_4$ (7.9 wt.\%) and Mg(AlH$_4$)$_2$ (6.9 wt.\%), which have initial hydrogen desorption temperatures of 453 K and 423 K, respectively \cite{Blanchard2004,Fichtner2004}.
However, these materials have low reaction enthalpies and therefore require extremely high pressures for the absorption of hydrogen.

Despite the unfavorable thermodynamics, it may be possible to change the decomposition temperature and pressure of the high capacity alanates by altering the
material composition. The possibility of mixing two different alkali metals (M and M$'$) to form a mixed alanate of M$_x$M$'_{1-x}$AlH$_4$ or M$_{3-x}$M$'_x$AlH$_6$ has been explored by recent
computational studies \cite{Lovvik2004,Arroyo2004}. These efforts suggest that the tetrahydride alanates may be relatively unstable with respect to cation mixing. An ab initio study by de Dompablo
and Ceder has shown Na$_{1-x}$Li$_x$AlH$_4$ favors phase separation at 0 K for $0<x<1$ \cite{Arroyo2004}. However, there are a number of stable mixed compounds predicted for the hexahydrides,
specifically the elpasolites M$_2$M$'$AlH$_6$ \cite{Lovvik2004}. The mixed alkali elpasolites do not revert, according to reaction \ref{naalh4}, to a mixed tetrahydride (M$_x$M$'_{1-x}$AlH$_4$) or a
mixture of monoalkali alanates (2MAlH$_4$ + M$'$AlH$_4$). Rather, the absorption/desorption of hydrogen occurs in a single step analogous to reaction \ref{na3alh6}:
\begin{equation}
\textup{M}_2\textup{M}'\textup{AlH}_6 \leftrightarrow 2\textup{MH} + \textup{M}'\textup{H} + \textup{Al} + \frac{3}{2}\textup{H}_2.
\label{reaction}
\end{equation}
Despite the number of predicted stable elpasolite phases, Na$_2$LiAlH$_6$ is the only composition that has been studied experimentally \cite{Bogdanovic1997,Zaluski1999}. The substitution of
a Li ion for a Na ion in Na$_3$AlH$_6$ to form Na$_2$LiAlH$_6$ lowers the equilibrium H$_2$ pressure by 20 bar at 484 K
\cite{Bogdanovic1997}. However, little else is known about the effects of alkali substitutions in the hexahydrides. In this study, the structural and thermodynamic properties of the elpasolites
(M$_2$M$'$AlH$_6$ where M$\neq$M$'$) were investigated and compared with those of the cryolites (M=M$'$) to better understand the thermodynamic changes induced by cation mixing. Thermodynamic values were
obtained by measuring pressure-composition isotherms and estimating the enthalpy and entropy of the reaction.

\section{Experimental}
Preparation of the bialkali alanates is accomplished using a number of conventional methods such as dry milling \cite{Huot1999} or wet chemical techniques
\cite{Claudy1982,Bogdanovic1997}. In this study, a tetrahydride alanate was mechanically alloyed with the appropriate alkali hydride(s) using either of the following two reactions:
\begin{equation}
\textup{M}'\textup{AlH}_4  +  2\textup{MH}  \rightarrow  \textup{M}_2\textup{M}'\textup{AlH}_6
\label{2prec}
\end{equation}
\begin{equation}
\textup{MAlH}_4  +  \textup{MH}  +  \textup{M}'\textup{H}  \rightarrow  \textup{M}_2\textup{M}'\textup{AlH}_6.
\label{3prec}
\end{equation}
Precursors of LiAlH$_4$ (95\%) and LiH (99.4\%) were purchased from Alfa Aesar, while NaH (95\%) and NaAlH$_4$ (90\%) were obtained from
Aldrich. KH was received from Fluka dispersed in mineral oil at a concentration of 35\%. The oil was removed via an octane wash under Ar gas and dried under vacuum. It should be noted that dry KH is
extremely pyrophoric and should only be handled in an inert atmosphere. Milling was performed in an Ar atmosphere with a Fritsch Pulverisette 6 planetary mill. The gas pressure and temperature were
monitored during milling to ensure that the majority of the hydrogen remained in the solid. The powders (1--2 g) were milled in a 250 mL stainless steel vial with seven 15 mm diameter stainless steel
balls (13.7 g) for up to 40 h at 200 rpm.  The Ti-doped material was prepared by milling 96 mol\% Na$_2$LiAlH$_6$ with 4 mol\% TiCl$_3$ (Aldrich) under the same conditions for 1 h.

Structural properties were determined using synchrotron x-ray diffraction (XRD). These experiments were performed on beamline X7A at the National Synchrotron Light Source of Brookhaven National
Laboratory. Prior to the diffraction study, each sample was annealed for approximately 20 h under high pressure H$_2$ gas (160 bar at 510 K for Na$_2$LiAlH$_6$, 40 bar at 570 K for K$_2$NaAlH$_6$ and
150 bar at 480 K for K$_2$LiAlH$_6$). After annealing, the powders were sieved using a 400 mesh screen (37 $\mu$m) and then sealed in 0.5 mm glass capillary tubes under Ar gas. The capillary was mounted on
the 2nd axis of the diffractometer.Ê A monochromatic beam was selected using a channel-cut Si(111) monochromator.Ê A gas-proportional position-sensitive detector (PSD), gated at the Kr-escape peak, was
employed for high-resolution ($\Delta d/d \sim 10^{-3}$) powder diffraction data measurements \cite{Smith1991}. The PSD was stepped in 0.25$^{\circ}$ intervals between 10$^{\circ}$ and 70$^{\circ}$ in
$2\theta$ with an increasing counting time at higher angle.Ê The capillary was spun during the measurement to provide better powder averaging.

Pressure-composition isotherms were measured using a Sievert's-type apparatus. The material was reacted in a stainless steel tube, which was heated using a resistive tape. The internal
sample temperature was monitored using a type K thermocouple. The absorption/desorption isotherms were measured by adding/removing an aliquot of H$_2$, allowing equilibrium to be reestablished, and
measuring the pressure change.

\section{Results and Discussion}

\subsection{Li$_2$NaAlH$_6$ and Li$_2$KAlH$_6$}
The preparation of Li$_2$NaAlH$_6$ was attempted using two different synthesis routes: reactions \ref{2prec} and \ref{3prec}. In both cases, a characterization of the reaction products using x-ray
diffraction showed the formation of a small amount Na$_2$LiAlH$_6$, but no other compositional changes. Similarly, attempts to synthesize Li$_2$KAlH$_6$ using reaction \ref{3prec} were
also unsuccessful. X-ray diffraction after milling revealed a reaction product consisting of KAlH$_4$ and LiH. These results suggest that Li$_2$NaAlH$_6$ and Li$_2$KAlH$_6$ are thermodynamically
less stable than other competing cryolite or elpasolite phases. Lovvik et.~al have predicted the enthalpy changes associated with mixing different alkali metals to form a bialkali alanate at 0 K using
density functional theory (DFT). The large, positive mixing enthalpies associated with Li$_2$NaAlH$_6$ and Li$_2$KAlH$_6$ (10.9 and 20.8 kJ/mol, respectively) are additional evidence that these
compounds are thermodynamically unstable
\cite{Lovvik2004}.

\subsection{Na$_2$LiAlH$_6$ and Na$_2$KAlH$_6$}

\begin{figure}
\includegraphics{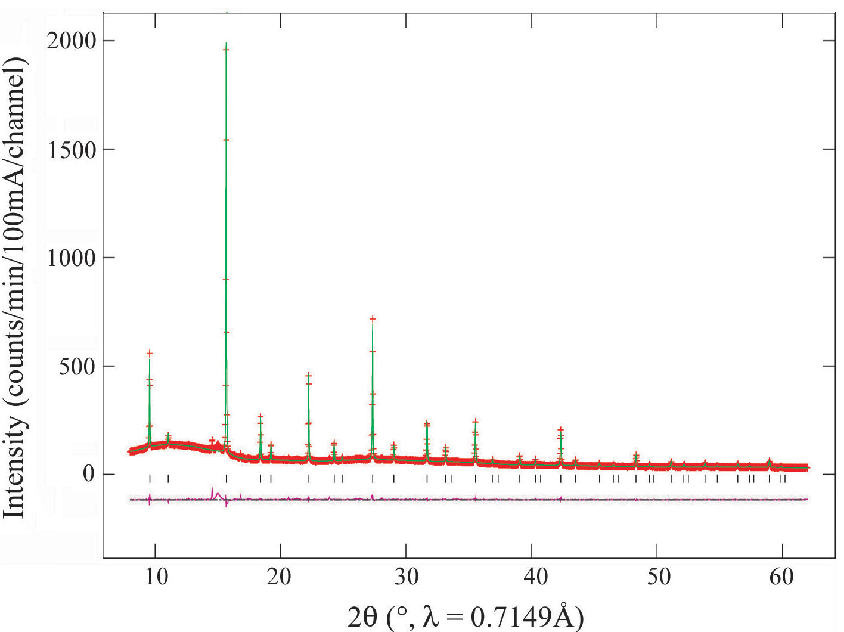} 
\caption{(Color online) Synchrotron x-ray powder diffraction pattern from Na$_2$LiAlH$_6$ (+) showing the Rietveld fit (solid), peak positions ($\shortmid$), and difference pattern (below). The
quality values for the Rietveld fit and weighted fit are Rp $ = 1.90$\% and wRp $ = 3.12$\%, respectively.}
\label{Na2LiAlH6}
\end{figure}

Na$_2$LiAlH$_6$ was prepared by both reaction \ref{2prec} and \ref{3prec}. A characterization of the material at different stages of alloying using XRD
showed that in reaction \ref{3prec} intermediate phases of NaAlH$_4$ and LiH are formed before the final product (Na$_2$LiAlH$_6$), while pathway \ref{2prec} leads to Na$_2$LiAlH$_6$ directly. The
structure of Na$_2$LiAlH$_6$ was determined by x-ray powder diffraction using a wavelength of 0.7149 $\textup{\AA}$ (Fig.~\ref{Na2LiAlH6}).
An elemental analysis, measured with an inductively coupled
plasma - mass spectrometer (ICP-MS), confirmed a near-stoichiometric composition of 2.02:1.06:1.00 for Na:Li:Al. A Rietveld structure refinement \cite{Rietveld,Young} using GSAS \cite{GSAS} was performed
with the constrained stoichiometry (Na$_2$LiAlH$_6$) and revealed the best fit in the $Fm\bar{3}m$ space group with a lattice constant of 7.4064(1) $\textup{\AA}$. This is in contrast with the room
temperature phases of the pure cryolites Li$_3$AlH$_6$ and Na$_3$AlH$_6$, which crystallizes in the R$\bar{3}$ \cite{Brinks2003} and $P2_1/n$ \cite{Ronnebro2000} space groups, respectively. The Rietveld fit
along with a difference plot are shown in Fig.~\ref{Na2LiAlH6}. The results of the Rietveld analysis, displayed in Table \ref{table1}, demonstrate that the Na ions are in a 12-fold coordination while the Li
ions occupy octahedral sites. 
\begin{table}
\caption{Interatomic distances and coordination numbers for Na$_2$LiAlH$_6$.}
\begin{ruledtabular}
\begin{tabular}{c c c }
{ Neighbors } & { Distance ($\textup{\AA}$) } & { Coordination } \\\hline
Na-H  & 2.6205(3) & 12  \\
Na-Al & 3.2071(1)  & 4   \\
Na-Li & 3.2071(1)  & 4   \\
Na-Na & 3.7032(1)  & 6   \\\hline
Li-H  & 1.952(8)    & 6   \\
Li-Na & 3.2071(1)  & 8   \\
Li-Al & 3.7032(1)  & ~6   
\label{table1}
\end{tabular}
\end{ruledtabular}
\end{table}
These values are consistent with the predictions of Lovvik et.~al \cite{Lovvik2004} and are
also in agreement with the space group and lattice constant reported by Claudy et.~al using conventional x-ray diffraction \cite{Claudy1982}.

\begin{figure} 
\includegraphics{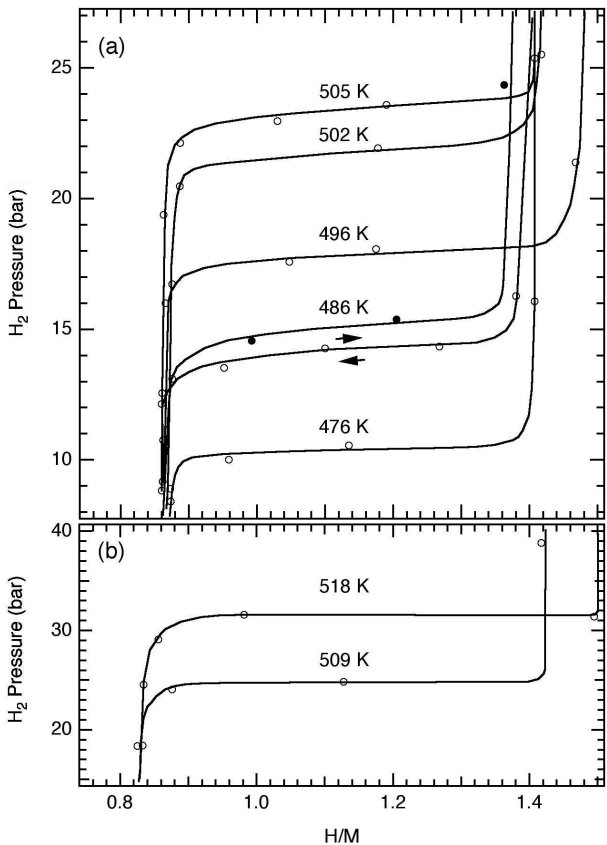} 
\caption{Pressure-composition absorption (\large{$\bullet$}\small) and desorption (\large{$\circ$}\small) isotherms for (a) Ti-doped Na$_2$LiAlH$_6$ and (b) uncatalyzed Na$_2$LiAlH$_6$. The arrows
indicate the direction of hydrogen transfer.}
\label{pci}
\end{figure}

The pressure-composition isotherms for catalyzed and uncatalyzed Na$_2$LiAlH$_6$ are displayed in Figs.~\ref{pci}a and \ref{pci}b, respectively.
Desorption
isotherms are shown at various temperatures between 476 K and 518 K. The isotherms exhibit a sharp transition upon desorption indicating the emergence of a new phase with little or no
solid solution region. Another sharp transition appears at the end of desorption, indicating the depletion of the Na$_2$LiAlH$_6$ phase. A complete absorption/desorption isotherm
taken at 486 K is also displayed in Fig.~\ref{pci}a. The complete isotherm exhibits little hysteresis as previously reported by Bogdanovic et.~al \cite{Bogdanovic1997}. The measured hydrogen
capacities for the catalyzed material are approximately 3.0 wt.\% (85\% of theoretical) for the first cycle (496 K) and 2.6 wt.\% (75\% of theoretical) for subsequent cycles. The hydrogen capacity for the
uncatalyzed material is 3.2 wt.\% (91\% of theoretical) for the first cycle (518 K) and 2.8 wt.\% (80\% of theoretical) for subsequent cycles. In both cases, the capacity loss after the first cycle is
attributed to incomplete absorption, which was confirmed by the presence of NaH, LiH and Al in XRD patterns after rehydriding.

\begin{figure} 
\includegraphics{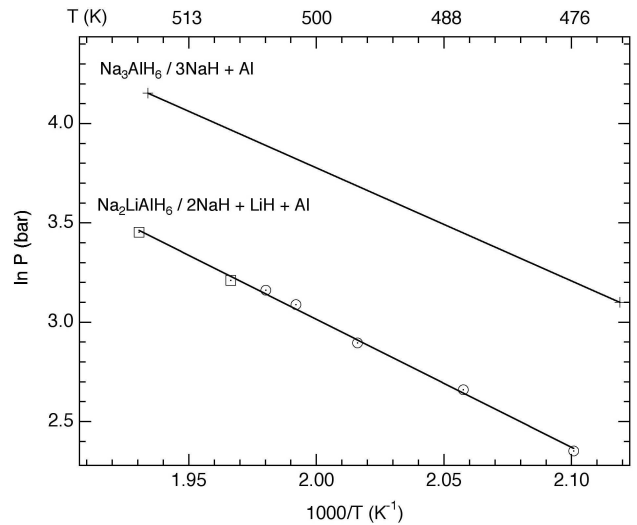} 
\caption{Van't Hoff plot for the reversible dissociation of Ti-doped Na$_3$AlH$_6$ (+) \cite{Bogdanovic2000} and Na$_2$LiAlH$_6$ (\large{$\circ$}\small-4 mol\% TiCl$_3$
\small{$\square$}\small-undoped).}
\label{hoff}
\end{figure}

The thermodynamic parameters of the decomposition reaction were calculated from the van't Hoff equation:
\begin{equation}
\textup{ln}P = \frac{1}{T}\Biggl(\frac{-\Delta H}{R} \Biggr) + \frac{\Delta S}{R},
\label{equation}
\end{equation}
where $P$ is the equilibrium H$_2$ pressure at temperature $T$ and $R$ is the universal gas constant. Using a plot of ln$P$ vs.~$1/T$, known as a van't Hoff plot, the enthalpy ($\Delta H$) and entropy
($\Delta S$) are determined from the slope and intercept, respectively. The van't Hoff plots for Ti-doped Na$_3$AlH$_6$ \cite{Bogdanovic2000} and Na$_2$LiAlH$_6$ (doped and undoped) are shown in
Fig.~\ref{hoff}.
The equilibrium pressure values were taken from the midpoint of the plateau in Fig.~\ref{pci}. The plateau pressure of Na$_3$AlH$_6$ is approximately 20 bar higher than that of
Na$_2$LiAlH$_6$ at 486 K and 30 bar higher at 518 K. The measured decomposition enthalpy, $\Delta H = 53.5 \pm 1.2$ kJ/mol H$_2$, is slightly more positive than the decomposition enthalpy of
Na$_3$AlH$_6$ (47 kJ/mol H$_2$ \cite{Bogdanovic2000}). The entropy change, $\Delta S = 132.1 \pm 2.4$ J/mol K of H$_2$, is approximately equivalent to the entropy associated with the formation of
molecular hydrogen ($S($H$_2)=130.7$ kJ/mol K).

The addition of a catalyst (4 mol\% TiCl$_3$) significantly improves the reaction kinetics. In the uncatalyzed state, Na$_2$LiAlH$_6$ begins slowly desorbing hydrogen at approximately 490 K, while the
catalyzed material exhibits a desorption temperature of less than 470 K. Similarly, the time required for the system to return to equilibrium is substantially decreased by the addition of a catalyst.
Equilibrium times ranged from 1--2 days for the Ti-doped samples to approximately 1 week for the undoped material.

Although the kinetics of reaction \ref{reaction} are clearly enhanced by the addition of Ti, the thermodynamics remain unaffected. This is demonstrated in Figure \ref{hoff}, which illustrates that the
data for the catalyzed and uncatalyzed material fall on the same line in the van't Hoff plot. The dopant has no measurable affect on the enthalpy or entropy  of the desorption reaction. This supports a
number of recent studies that suggest that the Ti acts as a true catalyst (possibly in the form of a Ti-Al alloy) \cite{Graetz2004,Felderhoff2004,ThomasGJ2002} and does not alter the reaction
thermodynamics.

The preparation of Na$_2$KAlH$_6$ was attempted using reaction \ref{3prec}. X-ray diffraction of the reaction product revealed phases of K$_2$NaAlH$_6$ and Na$_3$AlH$_6$. This is consistent with predicted
mixing enthalpies, which suggest that Na$_2$KAlH$_6$ is considerably less stable than K$_2$NaAlH$_6$ (by 40 kJ/mol) at 0 K \cite{Lovvik2004}. Therefore, Na$_2$KAlH$_6$ should favor phase separation into
equal parts K$_2$NaAlH$_6$ and Na$_3$AlH$_6$, as empirically observed.

\subsection{K$_2$LiAlH$_6$ and K$_2$NaAlH$_6$}
\begin{figure} 
\includegraphics{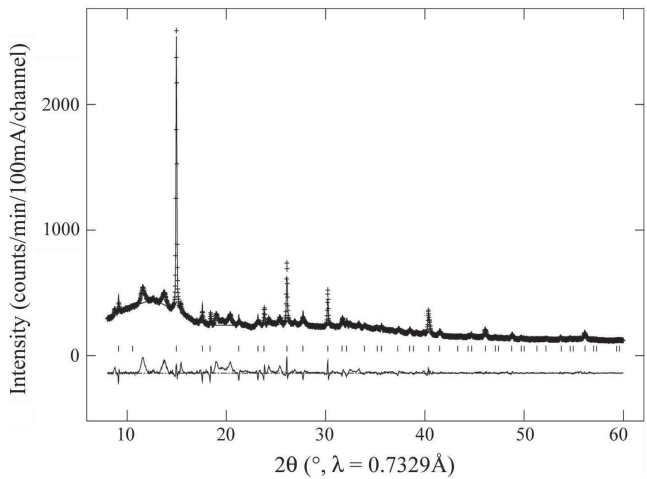} 
\caption{Synchrotron x-ray powder diffraction pattern from K$_2$LiAlH$_6$ (+) showing the profile fit (solid), peak positions ($\shortmid$), and difference pattern (below). The quality value for
the Lebail fit profile is wRp $ = 5.86$\%.}
\label{K2LiAlH6}
\end{figure}

A new alanate was prepared by mixing K and Li ions in reaction \ref{2prec} to form K$_2$LiAlH$_6$. An elemental analysis of the product (ICP-MS) revealed a near-stoichiometric composition of
2.04:1.10:1.00 for K:Li:Al. The powder diffraction results, using x-rays of wavelength 0.7329 $\textup{\AA}$, are shown in Fig.~\ref{K2LiAlH6}.
This pattern was not suitable for a Rietveld analysis due
to a number of additional broad peaks attributed to KH impurities. However, a full profile fit to the data \cite{LeBail1988} demonstrated the best fit in the $Fm\bar{3}m$ space group with a lattice
parameter of 7.9383(5) $\textup{\AA}$. As observed for Na$_2$LiAlH$_6$, the mixed potassium/lithium elpasolite crystallizes in a different space group than the pure cryolite phases
Li$_3$AlH$_6$ and K$_3$AlH$_6$, which have rhombohedral \cite{Brinks2003} and tetragonal \cite{Chini1966} symmetries, respectively. The full profile fit along with a difference pattern are displayed in
Fig.~\ref{K2LiAlH6}. The $Fm\bar{3}m$ space group is consistent with the structure predicted by DFT for K$_2$LiAlH$_6$ \cite{Lovvik2004}.

\begin{figure} 
\includegraphics{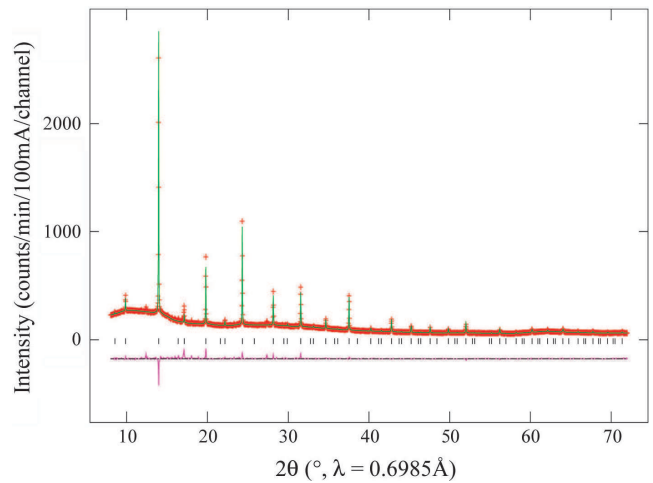} 
\caption{(Color online) Synchrotron x-ray powder diffraction pattern from K$_2$NaAlH$_6$ (+) showing the Rietveld fit (solid), peak positions ($\shortmid$), and difference pattern (below). The
quality values for the Rietveld fit and weighted fit are Rp $ = 2.05$\% and wRp $ = 3.08$\%, respectively.}
\label{K2NaAlH6}
\end{figure}

K$_2$NaAlH$_6$ was prepared by both synthesis routes (reaction \ref{2prec} and \ref{3prec}). The structural characterization was performed by XRD using a wavelength of 0.6985 $\textup{\AA}$
(Fig.~\ref{K2NaAlH6}).
An elemental analysis (ICP-MS) gave a composition of 1.54:0.97:1.00 for K:Na:Al. A small amount of NaH was detected in the diffraction pattern indicating that the product was
probably stoichiometric K$_2$NaAlH$_6$ with a NaH impurity ($\sim$20 mol\%). A sample prepared with excess KH ($\sim25$ mol\%) resulted in a similar diffraction pattern with the exception of small
amounts of KH impurities. Therefore, the Rietveld refinement using the data presented in Fig.~\ref{K2NaAlH6} was performed with a fixed stoichiometry of K$_2$NaAlH$_6$ (impurities were ignored).
The Rietveld fit, in the $Fm\bar{3}m$ space group, along with a difference plot are shown in Fig.~\ref{K2NaAlH6}. The measured lattice parameter is $a=8.1209(1)$ $\textup{\AA}$. The nearest-neighbor
distances and coordination numbers are listed in Table \ref{table2}.

\begin{table}
\caption{Interatomic distances and coordination numbers for K$_2$NaAlH$_6$.}
\begin{ruledtabular}
\begin{tabular}{c c c }
{ Neighbors } & { Distance ($\textup{\AA}$) } & { Coordination } \\ \hline
K-H   & 2.8841(3) & 12  \\
K-Al  & 3.5164(1)  & 4   \\
K-Na  & 3.5164(1)  & 4   \\
K-K   & 4.0604(1)  & 6   \\\hline
Na-H  & 2.3027(27)  & 6   \\
Na-K  & 3.5164(1)  & 8   \\
Na-Al & 4.0604(1)  & ~6 
\label{table2}
\end{tabular}
\end{ruledtabular}
\end{table}

The first, and only other reported synthesis of K$_2$NaAlH$_6$ used reaction \ref{2prec} in an organic medium under high pressure (25 kbar) H$_2$ gas. An XRD structural characterization of this
material suggested a monoclinic unit cell with dimensions $a=5.706$ $\textup{\AA}$, $b=5.707$ $\textup{\AA}$, $c=8.114$ $\textup{\AA}$ and $\beta=90.24^{\circ}$ \cite{Bastide1987}. The
discrepancy in the structural parameters may be attributed to the high pressure synthesis used in the earlier study, which may have distorted the unit cell from the cubic $Fm\bar{3}m$ structure.

\begin{figure} 
\includegraphics{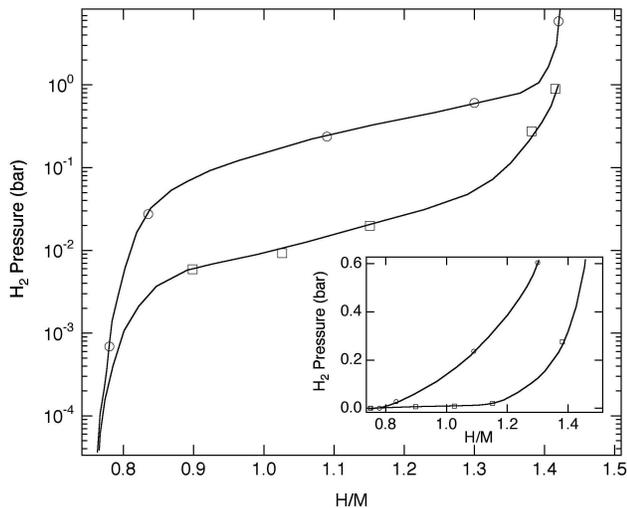} 
\caption{Pressure-composition absorption isotherms (no doping) for K$_2$LiAlH$_6$ (\large{{$\circ$}}\small) and K$_2$LiAlH$_6$ (\small{$\square$}) at 574 K plotted on a semi-log scale.
A plot of the data on a linear scale is displayed in the inset.}
\label{isotherms2}
\end{figure}

The pressure-composition absorption isotherms for K$_2$LiAlH$_6$ and K$_2$NaAlH$_6$ at 574 K are shown in Fig.~\ref{isotherms2}.
The initial desorption occurs at approximately 500 K for
K$_2$LiAlH$_6$ and 530 K for K$_2$NaAlH$_6$. At 574 K, the absorption reaction proceeds slowly in both samples. The time required for the system to reach equilibrium after the addition
of H$_2$ gas is around 280 h for K$_2$LiAlH$_6$ and greater than 1000 h for K$_2$NaAlH$_6$. Both isotherms exhibit a gradual increase in equilibrium pressure with respect to composition. The lack of a
clearly defined pressure plateau indicates that these materials do not exhibit a definitive two phase region at 574 K. It is likely that these data were collected at a temperature above the
critical temperature of the miscibility gap in the compositional phase diagram. The measured hydrogen storage capacity for K$_2$LiAlH$_6$ and K$_2$NaAlH$_6$ is 2.3 wt.\% and 2.0 wt.\%,
respectively. This capacity is $\sim90$\% of the theoretical value and the loss is attributed to an impurity of KOH in the KH precursor.

Due to the impractically slow reaction kinetics, it was not feasible to measure a series of desorption isotherms at different temperatures. Therefore, the reaction enthalpies for K$_2$LiAlH$_6$ and
K$_2$NaAlH$_6$ were approximated from equation \ref{equation}. The equilibrium pressures were taken from the midpoint of the isotherms shown in Fig.~\ref{isotherms2}. The entropy was estimated, $\Delta
S|_{T \rightarrow \infty} \approx 130.7$ kJ/mol K, using the change in entropy associated with the transition of hydrogen from an ordered solid to a gas. The estimated decomposition enthalpies (reaction
\ref{reaction}) for K$_2$LiAlH$_6$ and K$_2$NaAlH$_6$ are 82 kJ/mol H$_2$ and 97 kJ/mol H$_2$, respectively.

\begin{figure}
\includegraphics{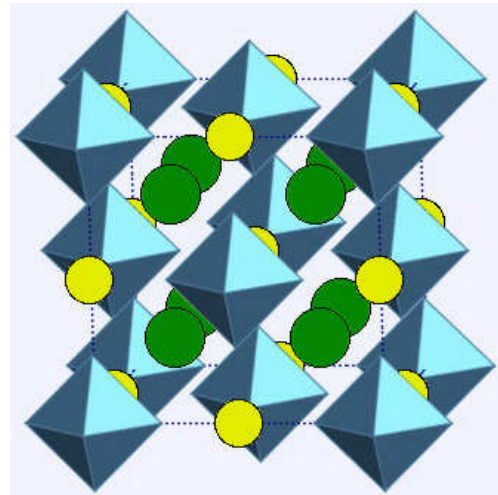}
\caption{(Color online) Structural diagram of M$_2$M$'$AlH$_6$ showing AlH$_6$ octahedra, M cations (large) and M$'$ cations (small).}
\label{elpasolite}
\end{figure}

A summary of the structural and thermodynamic data for the hexahydride alanates is presented in Table \ref{table}.
\begin{table*}
\caption{Structural and thermodynamic properties based upon experimental data for alanates of the form M$_2$M$'$AlH$_6$. The parameters listed include the structure (space group or symmetry),
lattice constants, decomposition enthalpy ($\Delta H$) and decomposition temperature ($T_\textup{d}$). The data for the cryolite phases were obtained from the references listed.}
\begin{ruledtabular}
\begin{tabular}{c c c c c c c }
{ M } & { M$'$ } & { Structure } & { Lattice Constants ($\textup{\AA}$) } & { $\Delta H$ (kJ/mol H$_2$) } &  { $T_\textup{d}$ (K) }  & { References } \\ \hline
   & Li & R$\bar{3}$   & $a=8.0712(1)$ $c=9.5130(2)$ &43.5      & 453 &\cite{Brinks2003}, \cite{Dymova1994}, \cite{Blanchard2004} \\
Li & Na & unstable     & -                                      & -    & - &  \\
   & K  & unstable     & -                                      & -    & - &  \\ \hline
   & Li & $Fm\bar{3}m$ & $a=7.4064(1)$                         & 53.5(12) & 490 &\\
Na & Na & $P2_1/n$     & $a=5.390(2)$ $b=5.514(2)$ & 47 & 473 &\cite{Ronnebro2000}, \cite{Bogdanovic2000}, \cite{Bogdanovic1997}\\&&&$c=7.725(3)$ $\beta=89.86(3)^{\circ}$&&&\\
   & K  & unstable     & -                                      & -    & -   & \\ \hline
   & Li & $Fm\bar{3}m$ & $a=7.9383(5)$                          & 82   & 500 & \\
K  & Na & $Fm\bar{3}m$ & $a=8.1209(1)$                         & 97   & 530 & \\
   & K  & tetragonal   & $a=8.445$ $b=8.584$                    & 135  & 593 &\cite{Chini1966}, \cite{Grochala2004}, \cite{Bastide1987}\\
\label{table}
\end{tabular}
\end{ruledtabular}
\end{table*}
The bialkali alanates have a face centered cubic structure in the $Fm\bar{3}m$ space group, similar to the high temperature (525 K) sodium alanate phase, $\beta$-Na$_3$AlH$_6$ \cite{Claudy1980}. In
these compounds ($\beta$-M$_2$M$'$AlH$_6$), the smaller M$'$ is octahedrally coordinated while M has a 12-fold coordination. A diagram of the general elpasolite structure is shown in Fig.
\ref{elpasolite}.
In the monoclinic ($P2_1/n$) polymorph, $\alpha$-M$_2$M$'$AlH$_6$, the M ion is coordinated by 8 atoms \cite{Bastide1981}. Ab initio calculations have predicted
$\alpha$-Na$_2$LiAlH$_6$ to be slightly more stable than the $\beta$-phase at 0 K \cite{Arroyo2004}. However, at 300 K the experimental data suggests that the
$\beta$ polymorph is preferred for all of the stable bialkali alanates. It is interesting to note that when two different alkali metals are present, the octahedral site is always occupied by the
smaller ion. The compounds Li$_2$NaAlH$_6$, Li$_2$KAlH$_6$ and Na$_2$KAlH$_6$ tend to phase separate, suggesting that the substitution of the smaller ion into the 12-fold coordinated site is
thermodynamically unfavorable. This trend also seems to apply to the "true" elpasolites, M$_2$M$'$AlF$_6$. Each of the stable bialkali aluminum hydrides (M, M$'$ = Li, Na or K) has a corresponding
fluoride compound with a similar lattice constant ($\pm 0.2$ $\textup{\AA}$) \cite{Holm,Graulich,Morss} in the $Fm\bar{3}m$ space group.  Analogous to the hydrides, there are no reported fluorides where
M$'$ is greater than M.

The decomposition temperature ($T_\textup{d}$) and enthalpy are also dependent upon the size of the alkali metals M and M$'$. The substitution of Li for Na to form Na$_2$LiAlH$_6$ increases $\Delta
H$ and $T_\textup{d}$ by 6.5 kJ/mol and 20 K, respectively. This case is an exception to the general rule that the decomposition temperature and enthalpy increase with the size of the alkali metal. This
trend is well documented for the monoalkali alanates, as shown  in Table \ref{table}. The tendency for greater hydride stability with a larger alkali metal also applies when only a partial
substitution of the metal is involved. For example in the K-M$'$ alanates, K$_2$NaAlH$_6$ is $14.6$ kJ/mol more stable than K$_2$LiAlH$_6$ and $38.5$ kJ/mol less stable than K$_2$KAlH$6$.

\section{Conclusion}
Novel elpasolite phases of the complex alanates were synthesized using conventional mechanical alloying techniques. Each of these compounds reversibly absorbs and desorbs hydrogen without a catalyst.
The addition of a catalyst to Na$_2$LiAlH$_6$ shows improved kinetics, but no change in the bulk thermodynamics. Structural analyses of the bialkali alanates demonstrates that the preferred space
group is $Fm\bar{3}m$ with the larger ion in a 12-fold coordination and the smaller occupying an octahedral site. In general, a smaller alkali ion (M or M$'$) reduces the reaction enthalpy of the hexahydride.
These results demonstrate that the temperature of hydrogen evolution and the equilibrium gas pressure can be tailored by appropriate substitutions of the alkali metals.

\section{Acknowledgements} This work, including research carried out at the NSLS (beamline X7A), was supported by an LDRD at Brookhaven and by the U.S. DOE under contract DE-AC02-98CH10886. The authors would
also like to thank Santanu Chaudhuri (BNL) for his insight on the structure and stability of the mixed alanates.

\end{document}